\documentclass[a4paper,oneside,12pt]{article}
\usepackage[utf8]{inputenc}
\usepackage[T1]{fontenc}
\usepackage{url}
\usepackage{xspace}
\usepackage{graphicx}
\graphicspath{{../Calculs/}{../Figures/}}
\usepackage{amssymb,amsmath}

\newcommand{\edonkey}{{\em eDonkey}\xspace}
\newcommand{\edonkeydsfr}{{\em edonkeyFR}\xspace}
\newcommand{\edonkeydsua}{{\em edonkeyUA}\xspace}
\newcommand{\ie}{{\em i.e.}\xspace}
\newcommand{\vs}{{\em vs}\xspace}

\newcommand{\kad}{{\em KAD}\xspace}

\newcommand{\paedocoeff}{{\em paedophile coefficient}\xspace}
\newcommand{\paedo}{{\em paedophile}\xspace}
\newcommand{\notpaedo}{{\em not paedophile}\xspace}
\newcommand{\mixed}{{\em mixed}\xspace}

\newcommand{\gnutella}{{\em Gnutella}\xspace}

\newcommand{\ipm}{\cite{QuantifyingIPM}\xspace}

\usepackage{comment}
\specialcomment{reference}{\begingroup\ttfamily\footnotesize}{\endgroup}
\excludecomment{reference}

\title{Comparing paedophile activity in different P2P systems}
\author{Rapha\"el Fournier, Thibault Cholez, Matthieu Latapy, Cl\'emence Magnien\\ 
Isabelle Chrisment, Ivan Daniloff, Olivier Festor}

\begin{document}
\maketitle
\thispagestyle{empty}


\begin{abstract}
  Peer-to-peer (P2P) systems are widely used to exchange content over the
Internet. Knowledge on paedophile activity in such networks remains limited
while it has important social consequences. Moreover, though there are different
P2P systems in use, previous academic works on this topic focused on one system
at a time and their results are not directly comparable.

We design a methodology for comparing \kad and \edonkey, two P2P systems
among the most prominent ones and with different anonymity levels. We monitor
two \edonkey servers and the \kad network during several days and record
hundreds of thousands of keyword-based queries. We detect paedophile-related
queries with a previously validated tool and we propose, for the first time, a
large-scale comparison of paedophile activity in two different P2P systems. We
conclude that there are significantly fewer paedophile queries in \kad than in
\edonkey (approximately 0.09\% \vs 0.25\%).

\end{abstract}

\section{Introduction}
\label{sec-introduction}

Paedophile activity is a crucial social issue and is often claimed to be prevalent in peer-to-peer (P2P) file-sharing systems \cite{saferInternet,cnet03riaa}. However, current knowledge on paedophile activity in these networks remains very limited. Recently, research works have been conducted to improve this situation by quantifying paedophile activity in \gnutella and \edonkey, two of the main P2P systems currently deployed \cite{hughes06is,QuantifyingIPM}. They respectively conclude that 1.6\% and 0.25\% of queries are of paedophile nature, but these numbers are not directly comparable as the authors use very different definitions and methods. Such comparisons are of high interest though, since differences in features of P2P systems, such as the level of anonymity they provide, may influence their appeal for paedophile users. 

In this paper, we perform for the first time such a comparison. We focus on the \kad and \edonkey P2P systems, which are both widely used and differ significantly in their architecture: while \edonkey relies on a few servers, \kad is fully distributed. This lack of centralisation may lead users to assume that \kad provides a much higher level of anonymity than \edonkey. Comparing the two systems sheds light on the influence of a distributed architecture on paedophile behavior and increases general knowledge on paedophile activity in P2P systems. 

Section~\ref{sec-data} describes our datasets and how we collected them. Section~\ref{sec-stats} presents our comparison of the amount of paedophile queries in \kad and \edonkey. Section~\ref{sec-ages} focuses on an important feature of paedophile activity: ages entered in queries. Finally, in Section~\ref{sec-inference} we infer the fraction of paedophile queries in \kad from the one in \edonkey, and Section~\ref{sec-conclu} presents our conclusions.

\section{Datasets and measurements}
\label{sec-data}

In order to compare paedophile activity in two different P2P systems, we first
need appropriate datasets, the collection of which is a challenge in itself. In
\kad and \edonkey, different kinds of measurements are possible.

In \edonkey, servers index files and providers for these files, and users submit keyword-based queries to servers to seek files of interest to them \cite{edonkey}. By monitoring such a server, one may collect all those queries \cite{tenweeks}. Here, we record all queries received by two of the largest \edonkey servers during a three-month period in 2010. The servers are located in different countries (France and Ukraine) and have different filtering policies: the French server indexes only non-copyrighted material, while the Ukrainian server openly indexes all submitted files. Monitoring two such different servers will allow us to compare them in order to know if server policy impacts our results.


To collect \kad data, we use the HAMACK monitoring architecture \cite{cholez2010}, which makes it possible to record the queries related to a given keyword by inserting distributed probes close to the keyword ID onto the \kad distributed hash table. We supervise 72 keywords, which we choose to span well the variety of search requests entered in the system, with a focus on paedophile activity: a set of 19 \paedo keywords (\textit{babyj, babyshivid, childlover, childporn, hussyfan, kidzilla, kingpass, mafiasex, pedo, pedofilia, pedofilo, pedoland, pedophile, pthc, ptsc, qqaazz, raygold, yamad, youngvideomodels}),  which are known to be directly and unambiguously related to paedophile activity in P2P networks; a set of 23 \mixed keywords (\textit{1yo, 2yo, 3yo, 4yo, 5yo, 6yo, 7yo, 8yo, 9yo, 10yo, 11yo, 12yo, 13yo, 14yo, 15yo, 16yo, boy, girl, mom, preteen, rape, sex, webcam}) frequently used in paedophile queries but also in other contexts (for instance,  {\em Nyo} stands for {\em N years old} and is used by both paedophile users and parents seeking games for children of this age); and a set of 30 \notpaedo keywords (\textit {avi, black, christina, christmas, day, doing, dvdrip, early, flowers, grosse, hot, house, housewives, live, love, madonna, man, new, nokia, pokemon, rar, remix, rock, saison, smallville, soundtrack, virtual, vista, windows, world}) used as a test group and {\em a priori} rarely used in paedophile queries. The sets of keywords result of the work on paedophile query detection presented in~\ipm. Notice that our set of keywords contains mainly common English words ({\em love,  early,  flowers}), but some are in other languages ({\em saison,  pedofilia}), and some are also brand names ({\em pokemon, nokia}).

Because of the differences in architectures of the two networks and of the measurement methodologies, we obtained very different datasets, which are not directly comparable: in \edonkey, we observe all queries from a subset of users whereas in \kad we only observe queries related to a given keyword, but from all users. In addition, based on various versions of \kad clients, the measurement tool only records the queries containing a monitored keyword placed in first position or being the longest in the query. As a consequence, with a short keyword such as {\em avi}, a name extension for video files, we almost only record queries in which it is the unique keyword, because otherwise it most likely is neither the longest nor the first word in any query. In order to obtain comparable datasets, we therefore limit our study to a subset of our datasets: the queries composed of exactly one word among the 72 keywords we monitor. 

As a result of this construction process, we obtain three datasets, which we call \edonkeydsfr, \edonkeydsua and \kad. They contain 241,152, 166,154 and 250,000 queries respectively, all consisting of a unique keyword from our list of 72 monitored keywords, which ensures that they are comparable. The server corresponding to the \edonkeydsfr dataset is located in France, while the one corresponding to \edonkeydsua is in Ukraine. Notice moreover that their large sizes make us confident in the reliability of our statistical results presented hereafter.

\section{Amount of paedophile queries in \edonkey versus \kad}
\label{sec-stats}

\begin{figure}[!h]
\centering
\includegraphics[width=.7\columnwidth]{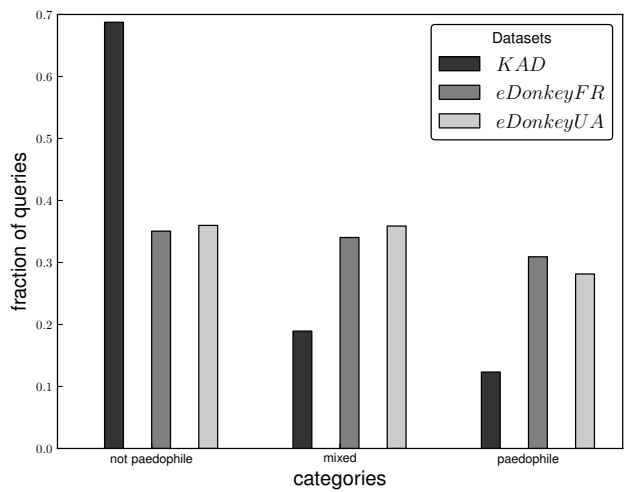}
\caption{
  Fraction of queries of each kind in our three datasets.
}
\label{fig-histocateg}
\end{figure}

The most straightforward way to compare the paedophile activity in different systems certainly is to compare the fraction of paedophile queries in each system. Figure~\ref{fig-histocateg} presents the fraction of queries for each category of keywords. This plot clearly shows that there are very distinct search behaviors in the two networks, since values obtained for the \paedo and \notpaedo categories significantly differ between \kad and the two \edonkey datasets. More surprisingly, the fraction of paedophile queries is significantly lower in \kad than in \edonkey which is in sharp contradiction with previous intuition, as \kad is assumed to provide a higher level of anonymity. The plot also shows that values obtained for the two \edonkey servers are similar, which indicates that very different filtering policies have no significant influence on the amount of paedophile queries.

\begin{figure}[!ht]
\centering
\includegraphics[width=.7\columnwidth]{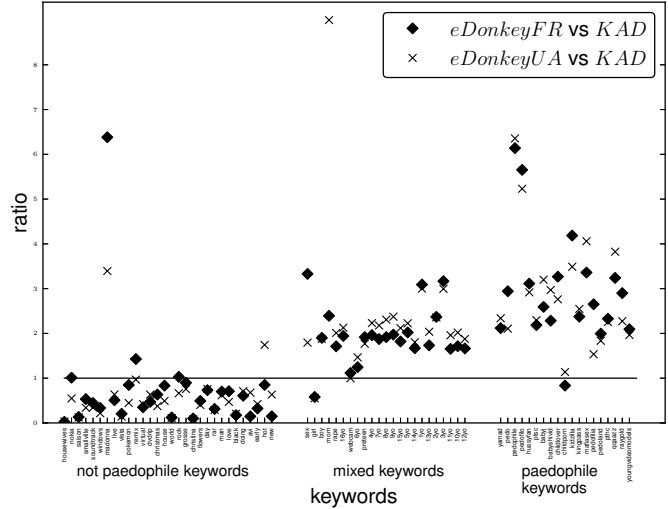}
\caption{
Ratio of keyword frequencies in \edonkey vs \kad. Keywords are ranked in increasing order of \paedocoeff. Points above the $y=1$ horizontal line indicates keywords more frequent in the corresponding \edonkey dataset; below the line keywords are more frequent in \kad. 
}
\label{fig-wordsOcc}
\end{figure}

In order to gain a more detailed insight on this phenomenon, we study the frequencies of each keyword separately in the three datasets. As we want to explore possible correlations between the paedophile nature of a keyword and its frequency, we need a way to quantify the paedophile nature of a keyword. To do so, we use the 28-week dataset and the paedophile query detection tool presented in~\ipm, which divides a dataset between \paedo and \notpaedo queries. We denote by $Q$ the whole dataset of queries, and by $Q(k)$ the set of queries containing a given keyword $k$. For each keyword $k$, we obtain $Q(k) = N(k) + P(k)$, where $N(k)$ and $P(k)$ are the subset of queries containing keyword $k$ and tagged as \notpaedo or \paedo, respectively. We then define the \paedocoeff $\pi(k)$ of keyword $k$ as: $\pi(k) = \frac{|P(k)|}{|Q(k)|}$. If all the queries with keyword $k$ are paedophile queries, $\pi(k) = 1$, and if none of them are, $\pi(k)=0$. All keywords in the \notpaedo category have a \paedocoeff below $0.006$. For keywords in the \mixed category, the \paedocoeff is above $0.01$ and below $0.4$. All \paedo keywords but one have a \paedocoeff above $0.85$. Finally, we plot in Figure~\ref{fig-wordsOcc} the ratios $\frac{f_{eDonkeyFR}(k)}{f_{kad}(k)}$ and $\frac{f_{eDonkeyUA}(k)}{f_{kad}(k)}$, where $f_s(k)$ denotes the frequency of queries composed of keyword $k$ in the dataset $s$, for each of our 72 keywords. We rank keywords on the horizontal axis in increasing order of \paedocoeff. The horizontal line represents {\em y = 1}, which enables a visual comparison of the values: if the point is below the line, then the keyword is more frequent in \kad, otherwise it is more frequent in the \edonkey dataset. 

This plot gives a clear evidence for a correlation between the paedophile nature of a keyword and its higher presence in \edonkey than in \kad. In addition, the frequencies in both \edonkey datasets are very similar for the vast majority of keywords.


We therefore conclude that anonymity is not the prevailing factor when paedophile users choose a network, since neither the decentralised architecture of \kad nor the different filtering policies increase the frequency of paedophile queries. Instead, the frequency of paedophile queries is even higher in \edonkey than in \kad. Finding an explanation for this unexpected phenomenon is still an open question. The higher technical skills required to use \kad may be part of the explanation. Users may also search content on \edonkey while protecting their privacy with other tools, such as Virtual Private Networks or TOR~\cite{torproject}. The fact that in \kad search requests are sent over UDP and cannot benefit from TOR anonymisation could explain the difference in the network usage. 


\section{Ages indicators in queries}
\label{sec-ages}
A way to gain more insight on observed paedophile activity is to study the distribution of age indicators in queries \cite{steel09child}. Notice that age indicators are sometimes used in other contexts than paedophile activity, especially when parents seek content suitable for children of a certain age. However, one can observe on Figure~\ref{fig-wordsOcc} that ages indicators have similar behavior to those obtained for the {\em paedophile} group, and are therefore closely related to the topic. 

\begin{figure}[!ht]
\centering
\includegraphics[width=.7\columnwidth]{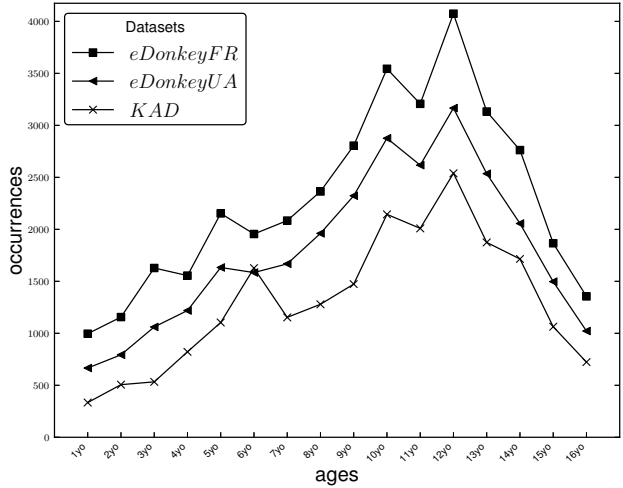}
\caption{
Distribution of age indicators in our three datasets.
}
\label{fig-ages}
\end{figure}

We plot the distribution of age indicators on Figure~\ref{fig-ages}: for each integer $n$ lower than 17, we plot the number of queries of the form {\em nyo} in each dataset ({\em yo} stands for {\em years old}). The three plots have similar shape, with mostly increasing values from 1 to 10, a little drop at 11, a peak at 12 and a fall from 13 to 16. These values for \kad are below the values for the \edonkey servers, which is due to the fact that this dataset is a bit smaller than others and that paedophile queries are rarer in it. The key point here is that the distributions are very similar in all three datasets. This indicates that, although the {\em amount} of paedophile activity varies between systems, its nature is similar, at least regarding ages.

\section{Quantifying paedophile activity in \kad}
\label{sec-inference}

In \ipm, the authors establish a method to quantify the fraction of paedophile queries in \edonkey. It relies on a tool able to accurately tag queries as paedophile or not, and on an estimate of the error rate of this tool. Such an approach cannot directly be applied to \kad though, as only a small (and biased) fraction of all queries may be observed in this system. We however show in this section how to derive the fraction of paedophile queries in \kad from the one in \edonkey.

\begin{figure}[!h]
\centering
\includegraphics[width=.3\columnwidth]{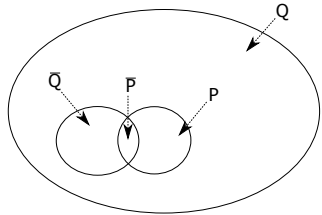}
\caption{
 The different sets of queries we define for each considered dataset.
}
\label{fig-sets}
\end{figure}

In a given system, \edonkey or \kad here, we consider different sets of queries and we denote by $Q$ the set of all queries, $P$ the subset of paedophile queries in $Q$, $\overline{Q}$ the subset of queries composed of one word among the 72 monitored keywords, $\overline{P}$ the subset of paedophile queries with one word, \ie consisting of one of the 19 monitored paedophile keywords (and so: $\overline{P}=\overline{Q}\cap P$). Figure~\ref{fig-sets} illustrates our notations.

In both our \edonkey measurements, $|P|$ and $|Q|$ may be directly estimated \ipm and one can then obtain the fraction $\frac{|P|}{|Q|}$ of paedophile queries in the dataset. We give the results for our two measurements in Table~\ref{table-inference}. On the contrary, in \kad, one may only estimate $|\overline{P}|$ and $|\overline{Q}|$. 

\begin{table}[!ht]
\centering
\begin{tabular}{|c|c|c|c|c|c|}
\hline
\rule[-1.55ex]{0cm}{4.5ex} dataset & $\frac{|P|}{|Q|}$ & $|\overline{P}|$ & $|\overline{Q}|$ & $\alpha$ & $\beta$ \\
\hline
\edonkeydsfr & $2.554 \cdot 10^{-3}$ & 74,557 & 241,152 & $1.431 \cdot 10^{-3}$ & 0.2502  \\
\hline
\edonkeydsua & $2.668 \cdot 10^{-3}$ & 46,763 & 166,154 & $1.538 \cdot 10^{-3}$  & 0.2251 \\ 
\hline
\kad & n/a & 30,821 & 250,000 & \mbox{n/a} & \mbox{n/a} \\
\hline
\end{tabular}
\caption{Main features of the three datasets.}
\label{table-inference}
\end{table}

However, we define 
$\alpha = \frac{|\overline{Q}|-|\overline{P}|}{|Q|-|P|}$ 
and 
$\beta = \frac{|\overline{P}|}{|P|}$
, which capture the probability for a non paedophile query, respectively paedophile, to make a query of one word among one of our monitored keywords. Given the definition of $\alpha$ and $\beta$, there is no {\em a priori} reason to assume that they have significantly different values between \edonkey and \kad. From the definitions of $\alpha$ and $\beta$, we have:
\begin{eqnarray*}
  \alpha = \frac{|\overline{Q}|-|\overline{P}|}{|Q|-|P|} & \implies & |Q| = |P| + \frac{|\overline{Q}|-|\overline{P}|}{\alpha} = \frac{\alpha |P| + |\overline{Q}|-|\overline{P}|}{\alpha}\\
  \beta = \frac{|\overline{P}|}{|P|} & \implies & |P| = \frac{|\overline{P}|}{\beta} 
\label{eq:xf}
\end{eqnarray*}
Then, the following expression holds: 
\begin{eqnarray}
  \frac{|P|}{|Q|} & = & \frac{|\overline{P}|}{\beta} \times \frac{\alpha}{\alpha |P| + |\overline{Q}|-|\overline{P}|} \nonumber\\
                  & = & \frac{\alpha |\overline{P}|}{\beta |\overline{Q}| + (\alpha - \beta) |\overline{P}|}  
\label{eq:xdef}
\end{eqnarray}

We now use expression (\ref{eq:xdef}) to infer the fraction of paedophile queries that were submitted in the \kad P2P network during our experiment. 
Using the values from Table~\ref{table-inference} and the average values of $\alpha$ and $\beta$ between our \edonkey datasets, we obtain: 

$$\frac{|P|}{|Q|} \thickapprox 0.087\% \pm 0.008$$
This value is of similar magnitude to the one of \edonkey (approx.
0.25\%) but close to three times lower. 

Notice that this estimation of $\frac{|P|}{|Q|}$ relies on the value of $\alpha$. One may wonder whether the choice of keywords from which we built $\overline{Q} \setminus \overline{P}$ has a significant impact on the estimated value of $\frac{|P|}{|Q|}$ in \kad. We check this as follows: we randomly select 1,000 subsets of 26 keywords out of the 53 keywords which compose the queries in $\overline{Q} \setminus \overline{P}$. We then compute, for each subset, the number of queries consisting of exactly one of those keywords and the resulting value of alpha. For \edonkeydsfr, we obtain an average value of $\overline{\alpha} = 0.000889 $ (minimum: 0.000256, maximum: 0.00153, and 90\% of the values in [0.000463;0.00133]). For \edonkeydsua, we obtain an average value of $\overline{\alpha} = 0.00105$ (minimum: 0.000352, maximum: 0.00172, and 90\% of the values in [0.00062;0.00148]). This means that we would obtain very similar results with 26 keywords only and so we may be confident in our estimate obtained with 53 keywords. 





%

\begin{reference}
  \begin{verbatim}

fournier@riazan:/data2/fournier/Loria_Data/ivan/
zcat peerbooter_lgr1_decoded_all.gz | python match7.py | grep ``^1'' | wc -l 
303611


ed2k\_1 :
- Tagged (T) pedo : 303611
- Queries (Q): 116694033
- coeff de correction infocom : (1-(15*192545/(207340*1000))/(1-(185/754))) = .98154136893071719562
- P barre : 74557
- Q barre : 241152
- P/Q = T/Q * corr = (303611/116694033)*.98154136893071719562 = .00255374459945543211 = 2.554.10^-3
- P = P/Q*Q = 303611*.98154136893071719562 = 298006.75656242397847938382
- alpha = Qbarre/Q = 241152/116694033 = .00206653239930442715 = 2.067.10^-3
- alpha = (Qbarre-Pbarre)/(Q-P) = (241152-74557)/(116694033-298007)  = .00143127738742558100
- beta = Pbarre/P = 74557/298006.75656242397847938382 = .25018560270254281741

ed2k\_2 :
- Tagged (T) pedo : 211670
- Queries (Q) : 77859581
- coeff de correction infocom : corr = (1-(15*192545/(207340*1000))/(1-(185/754))) = .98154136893071719562
- P barre : 46763 
- Q barre : 166154
- P/Q = T/Q * corr = (211670/77859581)*.98154136893071719562 = .00266843025473724175 = 2.668.10^-3
- P = P/Q*Q = 211670*.98154136893071719562 = 207762.86156156490879688540
- alpha = Qbarre/Q = 166154/77859581 = .00213402124524661903 = 2.134.10^-3
- alpha = (Qbarre-Pbarre)/(Q-P) = (166154-46763)/(77859581-207763) = .00153751712548442845
- beta = Pbarre/P = 46763/(211670*.98154136893071719562) = .22507872508361196603
- alpha/beta = (166154/77859581)/(46763/(211670*.98154136893071719562)) = .00948122149018693553

kad :
- P barre : 30821
- Q barre : 250000
- Pbarre/Qbarre = 30821/250000 = .12328400000000000000 
- P/Q = alpha/beta Pbarre/Qbarre :
    - .00825999728594064092*.12328400000000000000 = .00101832550539990597 
    - .00948122149018693553*.12328400000000000000 = .00116888291019620615




alpha1  0,0014312774
alpha2  0,0015375171
alpha0  0,0014843973 = (alpha1+alpha2)/2

beta1 0,2501856027
beta2 0,22507872
beta0 0,2376321614

Pbarre  30821
Qbarre  250000

P/Q
alpha1 beta1 0,0804%
alpha2 beta2 0,0960%
alpha0 beta0 0,0878%
alpha1 beta2 0,0008934062
alpha2 beta1 0,0008634363



  \end{verbatim}
\end{reference}


\section{Conclusion}
\label{sec-conclu}

We performed a first comparative study of two large-scale peer-to-peer networks, \kad and \edonkey, with regards to the queries related to child pornography. We designed a methodology to collect and process datasets allowing to compare them in a relevant manner. We obtained the counter-intuitive result that paedophile keywords are significantly more present in \edonkey than in \kad, despite the higher anonymity level it provides. On the contrary, our study of age indicators in queries showed that the nature of paedophile queries is similar in these systems. We finally established the first estimate of the fraction of paedophile queries in \kad. We obtained a value close to 0.09\%, which is of the same magnitude but significantly lower than in \edonkey (0.25\%). 

Our contributions open various directions for future work. 
In particular, our methodology may be applied to compare other systems, and our datasets may be used to perform either deeper analyses on paedophile activity or on general search engine behaviors. 

\section*{Acknowledgments}
This work is supported in part by the MAPAP SIP-2006-PP-221003 and ANR MAPE projects.

\bibliographystyle{plain}
\bibliography{biblio/bibliography,biblio/pedo,biblio/article_sigcomm}

\label{theend}
\end{document}